\documentclass[12pt]{article}
\begin{document}
\title{Black Hole Thermodynamics and Electromagnetism}
\author{B.G. Sidharth\\
International Institute for Applicable Mathematics \& Information Sciences\\
Hyderabad (India) \& Udine (Italy)\\
B.M. Birla Science Centre, Adarsh Nagar, Hyderabad - 500 063 (India)}
\date{}
\maketitle
\begin{abstract}
We show a strong parallel between the Hawking, Beckenstein black hole Thermodynamics and electromagnetism: When the gravitational coupling constant transform into the electromagnetic coupling constant, the Schwarzchild radius, the Beckenstein temperature, the Beckenstein decay time and the Planck mass transform to respectively the Compton wavelength, the Hagedorn temperature, the Compton time and a typical elementary particle mass. The reasons underlying this parallalism are then discussed in detail.
\end{abstract}  
A few decades ago, the work of Hawking, Beckenstein (and Unruh) and others brought out the connection between Thermodynamics, Black holes and Quantum Theory. We will now show that there is a striking parallel between gravitation and thermodynamical considerations on the one hand and electromagnetism on the other. We will then investigate the mechanism that leads to such a parallelism. Our starting point is the well known relation between the gravitational and electromagnetic coupling constants \cite{mwt}
\begin{equation}
\frac{Gm^2}{e^2} = \frac{1}{\sqrt{N}}\label{e1}
\end{equation}
In (\ref{e1})$m$ is the mass of a typical elementary particle which has been taken in the literature to be a pion and $N \sim 10^{80}$ is the well known number of elementary particles in the universe. Equation (\ref{e1}) is one of the Dirac large number relations and for this purpose it does not really matter if $m$ stands for the mass of a pion or a proton or an electron (Cf.\cite{mwt}). It may also be mentioned that (\ref{e1}) was considered to be a miraculous large number coincidence along with a few other such relations. However in recent years it has been shown that these relations can infact be deduced from the theory \cite{ijmpa,ijtp,psp,cu,sivaram1,sivaram2}. As such they are not empirical or accidental. In fact this theory correctly predicted an accelerating ever expanding universe with a small cosmological constant, as was subsequently observationally found.\\
In this scheme it was independently deduced (Cf.ref.\cite{ijmpa} et sequence) that
\begin{equation}
G = \frac{lc^2}{\sqrt{N}m} = \frac{l c^2 \tau}{m t} \equiv \Theta/t\label{e2}
\end{equation}
wherein we have used the relation $T = \sqrt{N}\tau$, this being an analogue of the Weyl-Eddington formula $R = \sqrt{N} l$. Both these again are deduced from theory rather than being ad hoc coincidences as earlier supposed. Equation (\ref{e2}) shows the dependence of $G$ on time, as in the Dirac and a few other cosmologies (Cf. above references) and leads to meaningful observational consequences including the otherwise unexplained anomalous accelerations of the Pioneer spacecrafts \cite{bgsnc,knmetric}. Equation (\ref{e2}) is just another form of (\ref{e1}). It was also pointed out \cite{fpl} that (\ref{e2}) shows up gravitation as an effect of electromagnetism spread over the $N$ particles of the universe.\\
We can also see this in the following way. What we are saying is that if $N \sim 1$ then $Gm^2$ can be replaced by $e^2$. Carrying this out on (\ref{e2}) we get
\begin{equation}
e^2 = lmc^2 \, \mbox{or}\, l = e^2/mc^2\label{e3}
\end{equation}
Apart from the fact that (\ref{e3}) is known to be correct, it also follows by a simple substitution of (\ref{e1}) in (\ref{e2}).\\
Let us now contrast the gravitational and electromagnetic aspects. It is known that for a Planck mass $m_P \sim 10^{-5}gm$, all the energy is gravitational and infact we have
$$Gm^2_P \sim e^2$$
For such a mass the Schwarzschild radius is the Planck length or Compton length for a Planck mass
\begin{equation}
\frac{Gm_P}{c^2} = l_P \sim \hbar/m_P c \sim 10^{-33}cm\label{e8}
\end{equation}
We can compare (\ref{e8}) with (\ref{e3}) which defines $l$ as what may be called the ``electromagnetic Schwarzschild'' radius viz., the Compton wavelength, when $e^2$ is seen as an analogue of $Gm^2$. To push these considerations further, we have from the theory of black hole thermodynamics \cite{ruffini} for any arbitrary mass $m$, first the Beckenstein temperature given by
\begin{equation}
T = \frac{\hbar c^3}{8\pi km G}\label{e6}
\end{equation}
Equation (\ref{e6}) gives the thermodynamic temperature of a Planck mass black hole. Further, as is well known,
\begin{equation}
\frac{dm}{dt} = - \frac{\beta}{m^2},\label{e9}
\end{equation}
where $\beta$ is given by
$$\beta = \frac{\hbar c^4}{(30.8)^3 \pi G^2}$$
This leads back to the usual black hole life time given by
\begin{equation}
t = \frac{1}{3\beta} m^3 = 8.4 \times 10^{-24} m^3 secs\label{e7}
\end{equation}
Let us now factor in the time variation of $G$ into (\ref{e9}). Essentially we use (\ref{e2}). Equation (\ref{e9}) now becomes
$$m^2 dm = -B \, \mu^{-2} t^2 dt, B \equiv \frac{\hbar c^4}{\lambda^3 \pi}, \,  \mu \equiv \frac{l c^2 \tau}{m}, \lambda^3 = (30.8)^3 \pi$$
Whence on integration we get
\begin{equation}
m = \frac{\hbar}{\lambda \pi^{1/3}} \left\{ \frac{1}{l^6}\right\}^{1/3} t \, = \frac{\hbar}{\lambda \pi^{1/3}} \frac{1}{l^2} t\label{e10}
\end{equation}
If we use the pion mass, $m$ in (\ref{e10}), we get for $t$, the pion Compton time.\\
In fact if we use (\ref{e2}) in (\ref{e6}) with the appropriate expression for $\Theta$, we get
$$kT = \frac{mc^3t}{l}$$
Using for $t$, the pion Compton time, we get
\begin{equation}
kT = mc^2\label{e11}
\end{equation}
Equation (\ref{e11}) is the well known relation expressing the Hagedorn temperature of elementary particles \cite{sivaram1}. It is an analogue of (\ref{e6}). Alternatively if we carry out the substitution $Gm^2 \to e^2$ in (\ref{e6}) in the above, we recover (\ref{e11}). Similarly instead of (\ref{e9}) we will get, with such a substitution,
$$\frac{dm}{dt} = - \frac{\hbar c^4}{\lambda^3 e^4} m^2,$$
Whence we get for the life time
\begin{equation}
\frac{\hbar c^4}{\lambda^3 e^4} t = \frac{1}{m}\label{e12}
\end{equation}
For an elementary particle, (\ref{e10}) and (\ref{e12}) are the same. Further from (\ref{e12}) we get, for the pion,
$$t \sim 10^{-23} secs,$$
which is again the pion Compton time. So the Compton time shows up as an ``electromagnetic Beckenstein radiation life time.''\\
Thus for elementary particles, working within the context of gravitational theory, but with either a time varying Gravitational constant being taken into consideration as in steps leading to (\ref{e10}) or alternatively a scaled up coupling constant with the rule $Gm^2 \to e^2$, we get the meaningful relations (\ref{e3}) and (\ref{e10}) and (\ref{e11}) giving the Compton length and Compton time as also the Hagedorn temperature as the analogues of the Schwarzschild radius, radiation life time and black hole temperature obtained with the usual gravitational coupling constant. The converse holds good for $e^2 \to Gm^2$. The parallel is complete. The analogy can be summarized as above.\\ \\
\begin{table}
\begin{tabular}{|l|l|} \hline
Coupling : $Gm^2$  & Coupling : $e^2$ \\
Schwarzchild radius & Compton wavelength \\
Beckenstein Temperature & Hagedorn temperature \\
Beckenstein decay time  & Compton time \\ 
Planck mass $m_P$  & Planck mass $m_\pi$ \\ \hline
\end{tabular}
\end{table}
{\large {\bf REMARKS}}\\ \\
1. It thus appears that gravitation plays out in the Beckenstein decay time while electromagnetism plays out in the Compton time, though through the same set of equations. One way of describing this phenomena is through (\ref{e1}), which no longer needs to be considered as a coincidence. Let us introduce two time scales (Cf. also \cite{csf}) $t_G$ and $t_e$. If we remember that the interaction is represented by the Hamiltonian,
$$H (t) = \imath \hbar \frac{d}{dt},$$
then we have for the two time scales
$$H (\equiv Gm^2) = \imath \hbar \frac{d}{dt_G}, \, H(\equiv e^2) = \imath \hbar \frac{d}{dt_e}$$
Using (\ref{e1}) we now get
\begin{equation}
H (t_G) = \frac{1}{\sqrt{N}} H(t_e)\label{e13}
\end{equation}
This is the bridge between the two interactions. Infact using this bridge in (\ref{e9}), viz.,
$$\frac{dm}{dt} = - \frac{\beta}{m^2}$$
We return to the time dependent gravitational constant as described after equation (\ref{e9}) and as in the Dirac cosmology and as described in detail in references \cite{ijmpa,ijtp,cu}. The subsequent equations then get modified.\\
Another way of seeing this is by realizing that our time is the elementary particle or ``electromagnetic'' time $t_e$. If we go to the gravitational ``time'' $t_G$ then we have to replace in (\ref{e10}) $t$ by $\sqrt{N}t$. This has an effect of replacing $l$ in (\ref{e10}) by $l_P$ the Planck length. In this case we recover the Planck time as the life time for a Planck mass $m_P$ on the one hand, and on the other, if the mass were that of the universe $\sim 10^{55}gms$, then we recover the life time as $10^{17}secs$, that is the age of the universe.\\ 
2. It is interesting that we can pursue the gravito thermodynamic link with electromagnetism further. Infact if we start with the Langevin equation in a viscous medium \cite{rief,balescu} then as the viscosity becomes vanishingly small, it turns out that the Brownian particle moves according to Newton's first law as if there were no force acting on it, that is with a constant velocity. Moreover this constant velocity is given by (Cf.refs.\cite{rief,balescu}), for any mass $m$,
\begin{equation}
c^2 \equiv \langle v^2 \rangle = \frac{kT}{m}\label{e1a}
\end{equation}
We would like to study the case where $m \to 0$. Then so too should $T$ for a meaningful limit. More realistically, let us consider (\ref{e1a}) with minimal values of $T$ and $m$, in the real world. Infact we have the thermodynamic Beckenstein formula \cite{ruffini} (\ref{e6})
We consider in (\ref{e6}) the entire universe so that the mass $M$ is $\sim 10^{55}gms$. Substitution in (\ref{e6}) gives
\begin{equation}
T = \frac{10^4}{10^{32}} \sim 10^{-28^{\circ}}K\label{e3a} 
\end{equation}
We next consider in (\ref{e1a}),m to be the smallest possible mass allowed thermodynamically. From thermodynamical considerations, Landsberg has shown that this is given by \cite{land}
\begin{equation}
m \sim 10^{-65} gms\label{e4a}
\end{equation}
The same equation (\ref{e4a}) can also be obtained from a different point of view, namely the Planck scale underpinning for the universe in modern Quantum Gravity approaches \cite{psp}. Substitution of (\ref{e3a}) and (\ref{e4a}) in (\ref{e1a}) gives
$$
\langle v^2 \rangle = \frac{kT}{m} = \frac{10^{-16} \times 10^{-28}}{10^{-65}} = 10^{21}, i.e.$$
\begin{equation}
v = c \, (cm/sec)\label{e5a}
\end{equation}
We can see from (\ref{e1a}) and (\ref{e5a}) that the velocity $c$ is exactly the velocity of light!\\
The question is, for how long such a particle with vanishingly small inertial mass can maintain the constant velocity $c$, that is the velocity of light. Infact the energy uncertainty $mc^2$ is associated with the lifetime $\sim \hbar/mc^2$, the Compton time. The Compton time for a particle with mass given by (\ref{e4a}) as can be easily checked is, $10^{17}secs$, which is the age of the universe! We should recover the same result from (\ref{e10}) or (\ref{e12}), and indeed we do!\\
3. If we consider the gravitational self energy of an elementary particle $m$, then this is given by
$$E_s = \frac{Gm^2}{l}$$
It is easy to verify that the Heisenberg Uncertainty Principle gives
\begin{equation}
E_s t = \hbar\label{e16}
\end{equation}
In (\ref{e16}) $t$ is the age of the universe \cite{sivaram1}. On the other hand the gravitational self energy of all the $N$ elementary particles in the universe is given by
\begin{equation}
E_u = \frac{NGm^2}{l}\label{e5}
\end{equation}
Such a gravitational energy has a low Beckenstein temperature which for mass $m$ is given by (\ref{e6}) or (\ref{e7}).\\
Substitution of (\ref{e5}) shows that the life time $t \sim 10^{17}secs$, that is the life time of the universe itself.\\
4. In the introduction, reference was made to the work of Unruh. He discovered that an uniformly accelerating object shows an event horizon and the consequent Beckenstein temperature is given by \cite{ciami},
\begin{equation}
T = \hbar a / 2 \pi kc\label{e18}
\end{equation}
It is interesting to note that in (\ref{e18}) if we replace the uniform acceleration $a$ by the uniform acceleration of rotation,
$$a = v^2/r$$
and further take $v$ to be $c$ and $r$ to be successively $l$, the Compton length or $R$ the Schwarzchild radius of a black hole, then we recover respectively the Hagedorn temperature for an elementary particle and the Beckenstein temperature (\ref{e6}) for a black hole.\\
There is another interesting fact. Let us, in (\ref{e18}), take $a$ to be $\Lambda R$, where $\Lambda$ is the cosmological constant and $R$ the radius of the universe. Then, using the value (\ref{e3a}) for $T$ (the Beckenstein temperature for the mass $M$ of the universe), we get the correct value
$$\Lambda \sim H^2 \sim c^2 / R^2$$
where $H$ is the Hubble constant. This should not be surprising because the acceleration $a$ in (\ref{e18}) is in the context of the universe, caused by the vaccum or dark energy.\\ 
On the other hand if we apply the Unruh formula (\ref{e18}) to the viscous zero point field, as in considerations of point 2, we recover the Hagedorn formula.\\
5. We note that the relation
$$\frac{\lambda^3 e^4}{\hbar^2 c^2} \sim 1$$
which follows from (\ref{e12}) on using the expression for the Compton time for $t$ gives an estimate for the fine structure constant $\sim 1/150$.\\
6. We have mentioned that gravitation is a form of weak electromagnetism. One way in which this can be understood is by realizing that the universe is by and large electrically neutral, because the atoms consist of an equal number of positive and negative charges. Strictly speaking these atoms are therefore electrical dipoles.\\
With this background let us consider the following simple model of an electrically neutral atom which nevertheless has a dipole effect. Infact as is well known from elementary electrostatics the potential energy at a distance $r$ due to the dipole is given by
\begin{equation}
\phi = \frac{\mu}{r^2}\label{ea1}
\end{equation}
where $\mu = eL, L \sim 10^{-8}cm \sim 10^3 l = \beta l,$ $e$ being the electric charge of the electron for simplicity and $l$ being the electron Compton wavelength. (There is a factor $cos \Theta$ with $\mu$, but on an integration over all directions, this becomes an irrelevant constant factor $4 \pi$.)\\
Due to (\ref{ea1}), the potential energy of a proton $p$ (which approximates an atom in terms of mass) at the distance $r$ (much greater than $L$) is given by
\begin{equation}
\frac{e^2L}{r^2}\label{ea2}
\end{equation}
As there are $N \sim 10^{80}$ atoms in the universe, the net potential energy of a proton due to all the dipoles is given by
\begin{equation}
\frac{Ne^2L}{r^2}\label{ea3}
\end{equation}
In (\ref{ea3}) we use the fact that the predominant effect comes from the distant atoms which are at a distance $\sim r$, the radius of the universe.\\
We now use the well known Eddington formula encountered earlier,
\begin{equation}
r \sim \sqrt{N} l\label{ea4}
\end{equation}
$r$ being of the order of the dimension of the universe. If we introduce (\ref{ea4}) in (\ref{ea3}) we get, as the energy $E$ of the proton under consideration
\begin{equation}
E = \frac{\sqrt{N}e^2\beta}{r}\label{ea5}
\end{equation}
Let us now consider the gravitational potential energy $E'$ of the proton $p$ due to all the other $N$ atoms in the universe. This is given by
\begin{equation}
E' = \frac{GMm}{r}\label{ea6}
\end{equation}
where $m$ is the proton mass and $M$ is the mass of the universe.\\
Comparing (\ref{ea5}) and (\ref{ea6}), not only is $E$ equal to $E'$, but remembering that $M = Nm$, we get back equation (\ref{e1}),
$$\frac{e^2}{Gm^2} = \frac{1}{\sqrt{N}}$$
7. In earlier communications, we had considered the Modified Uncertainty Principle which is there in Quantum Gravity approaches, including that of the author and also in Quantum Super String theory \cite{mup1,mup2,mup3}. It was argued that the Modified Uncertainty Principle is
\begin{equation}
\Delta x = \frac{\hbar}{\Delta p} + L^2 \frac{\Delta p}{\hbar}\label{e19}
\end{equation}
with a similar equation for the time coordinate, where $L$ stands for the Planck length or more generally the Compton length.\\
It was argued that while the first term on the right side of (\ref{e19}) gives the usual Uncertainty Principle of Heisenberg, the extra term is the contribution of gravitational effects. We can easily see that an application of the second term to the time coordinate leads us back to point 1 of this section, if we recognize that $\Delta E = \sqrt{N} mc^2, \sqrt{N}$ being the fluctuation or uncertainty in the number of particles in the universe. 

\end{document}